\documentclass{aa}  
\usepackage{graphicx}
\usepackage{txfonts}
\usepackage{lipsum}
\usepackage{booktabs}
\usepackage{tabularx}
\usepackage{enumitem}
\usepackage{float}
\newcolumntype{Y}{>{\centering\arraybackslash}X}
\usepackage{graphicx}
\usepackage{afterpage}
\usepackage{lscape}
\usepackage{longtable}
\usepackage{caption}
\usepackage{subcaption}
\usepackage{xcolor}
\usepackage{appendix}
\begin{document}

\title{Absence of radio-bright dominance in a near-infrared selected sample of red quasars\thanks{Table~\ref{tab:app} is only available in electronic form at the CDS via anonymous ftp to cdsarc.u-strasbg.fr (130.79.128.5) or via http://cdsweb.u-strasbg.fr/cgi-bin/qcat?J/A+A/}}

\author{S.~Vejlgaard\inst{1,2}
\and
J.~P.~U.~Fynbo\inst{1,2}
\and
K.~E.~Heintz\inst{1,2}
\and
J.~K.~Krogager\inst{3}
\and
P.~Møller\inst{4,2}
\and
S.~J.~Geier \inst{5,6}
\and
L.~Christensen\inst{1,2}
\and
G.~Ma\inst{1,2}
}

\institute{Cosmic Dawn Center (DAWN), \\
\email{simone.vejlgaard@nbi.ku.dk}
\and
Niels Bohr Institute, University of Copenhagen, Jagtvej 155, 2200 Copenhagen N, Denmark
\and
Centre de Recherche Astrophysique de Lyon, Univ. Claude Bernard Lyon 1, 9 Av. Charles Andre, 69230 St Genis Laval, France
\and 
European Southern Observatory, Karl-Schwarzschild-Straße 2, D-85748 Garching bei München, Germany
and 
Instituto de Astrof{\'i}sica de Canarias, V{\'i}a L{\'a}ctea, s/n, 38205, La Laguna, Tenerife, Spain
\and
Gran Telescopio Canarias (GRANTECAN), Cuesta de San Jos\'e s/n, E-38712, Bre\~{n}a Baja, La Palma, Spain
}

\date{Received --; accepted --}

\abstract{The dichotomy between red and blue quasars is still an open question. It is debated whether
red quasars are simply blue quasars that are observed at certain inclination angles or
if they provide insight into a transitional phase in the evolution of quasars.}
{We investigate the relation between quasar colors and radio-detected fraction because radio observations of quasars provide a powerful tool in
distinguishing between quasar models.}{We present the eHAQ+GAIA23 sample, which
contains quasars from the High A(V) Quasar (HAQ) Survey, the Extended High A(V)
Quasar (eHAQ) Survey, and the Gaia quasar survey. All quasars in this sample have been found using a
near-infrared color selection of target candidates that have otherwise been
missed by the Sloan Digital Sky Survey (SDSS). We implemented a redshift-dependent color cut in $g^*-i^*$ to select red quasars in the sample and divided
them into redshift bins, while using a nearest-neighbors algorithm to control
for luminosity and redshift differences between our red quasar sample and a
selected blue sample from the SDSS. Within each bin, we cross-matched the quasars
to the Faint Images of the Radio Sky at Twenty centimeters (FIRST) survey and
determined the radio-detection fraction.}{For redshifts 0.8 $<$ z $\leq$ 1.5,
the red and blue quasars have a radio-detection fraction of
0.153$^{+0.037}_{-0.032}$ and 0.132$^{+0.034}_{-0.030}$, respectively. The red
and blue quasars with redshifts 1.5 $<$ z $\leq$ 2.4 have radio-detection
fractions of 0.059$^{+0.019}_{-0.016}$ and 0.060$^{+0.019}_{-0.016}$,
respectively, and the red and blue quasars with redshifts z $>$ 2.4 have
radio-detection fractions of 0.029$^{+0.017}_{-0.012}$ and
0.058$^{+0.024}_{-0.019}$, respectively. For the WISE color-selected red
quasars, we find a radio-detection fraction of 0.160$^{+0.038}_{-0.034}$ for
redshifts 0.8 $<$ z $\leq$1.5, 0.063$^{+0.020}_{-0.017}$ for redshifts 1.5 $<$
z $\leq$ 2.4, and 0.051$^{+0.030}_{-0.022}$ for redshifts z $>$ 2.4. In other
words, we find similar radio-detection fractions for red and blue quasars
within $<1\sigma$ uncertainty, independent of redshift. This disagrees
with what has been found in the literature for red quasars in SDSS. It should
be noted that the fraction of broad absorption line (BAL) quasars in red SDSS quasars is about five times lower. BAL quasars have been observed to be
more frequently radio quiet than other quasars, therefore the difference in BAL fractions
could explain the difference in radio-detection fraction.}{The dusty torus of a
quasar is transparent to radio emission. When we do not observe a difference
between red and blue quasars, it leads us to argue that orientation is the main
cause of quasar redness. Moreover, the observed higher proportion of BAL
quasars in our dataset relative to the SDSS sample, along with the higher
rate of radio detections, indicates an association of the redness of
quasars and the inherent BAL fraction within the overall quasar population.
This correlation suggests that the redness of quasars is intertwined with the
inherent occurrence of BAL quasars within the entire population of quasars. In
other words, the question why some quasars appear red or exhibit BAL
characteristics might not be isolated; it could be directly related to the
overall prevalence of BAL quasars in the quasar population. This finding
highlights the need to explore the underlying factors contributing to both the
redness and the frequency of BAL quasars, as they appear to be interconnected
phenomena.}


\maketitle

\section{Introduction}
\label{sec:intro}
Even though the study of quasi-stellar objects (QSOs, or quasars) began more
than half a century ago \citep{Schmidt1963,Greenstein1964,Sandage1965}, it
remains a field in development. With extremely high bolometric luminosities (up
to $\sim10^{47}$ erg~s$^{-1}$; see, e.g., \citeauthor{Onken2022} 2022), quasars
are not only the most powerful class of active galactic nuclei (AGN), but also
some of the most luminous and distant objects known in the observable Universe
\citep{Wu2015, Wang2021}. Quasars are powered by the rapid accretion of matter
onto a supermassive black hole (SMBH) with a possible mass ranging up to
$10^{10}$ M$_\odot$ \citep[see also][]{Rees1984}. According to the unified
model for AGN and quasars presented, for instance, by \citet{Antonucci1993} and
\citet{Urry1995}, a dust torus absorbs and reemits photons from the accretion
disk with a dependence on the observed viewing angle. In addition to the broad-
and narrow-line regions producing the emission lines used to classify the
quasar, the unified model also describes quasar outflows. These outflows can
take the form of relativistic and collimated radio jets or more extended winds
originating from the accretion disk. In a subset of quasars called broad
absorption line (BAL) quasars, the strong winds create broad blueshifted
absorption lines in the UV part of the quasar spectrum
\citep{Foltz1987,Weymann1991}. \newline \indent A typical quasar spectrum is
characterized by a rest-frame blue or UV power-law continuum on which broad
emission lines are superimposed. However, a number of studies have already
confirmed the existence of a quasar population with a much redder continuum,
which is simply referred to as "red" quasars \citep{Webster1995,Benn1998,Warren2000,
Gregg2002,Hopkins2004, Glikman2007,Fynbo2013}. Our understanding of this
particular subset of quasars is still far from complete. While many studies
attribute the red optical and infrared (IR) colors to dust obscuration
\citep{Sanders1988a, Sanders1988b}, other plausible explanations include
starlight contamination from the quasar host galaxy \citep{Serjeant1996} and
differences in the accretion rates \citep{Young2008}. From the attempts to
explain the observational differences, two main quasar redness paradigms have
emerged: The orientation model, and the evolutionary model. The orientation
model claims that any observed differences between red and blue quasars are due
to the observer's viewing angle with respect to the dusty torus, such that
inclinations closer to the equatorial plane of the torus correspond to a redder
quasar classification \citep{Antonucci1993}. According to the orientation
model, the differentiation between red and blue quasars is thus not
grounded in physical circumstance, but solely a consequence of relative positioning with respect to
either of these objects. The evolutionary model claims that an intrinsic
physical evolution is the origin of the quasar redness, such that the red quasar
population represents a transitional phase between an early highly
dust-obscured star-forming stage and the blue quasar stage \citep{Sanders1988a,
Sanders1988b, Hopkins2006, Hopkins2008, Alexander2012, Glikman2012}. Within
this framework, the host galaxies of red quasars are thought to have undergone
a major galaxy merger, which induces either strong winds or jets in the central
red quasar, with the ability to dissolve their surrounding dust cocoons. This
quenches the host galaxy star formation, and the underlying unobscured blue
quasar is revealed. 

The fact that the optical colors of dust-reddened quasars resemble those of
low-mass stars \citep{Richards2002, Richards2003} motivated the use of radio
selection to build red quasar samples without stellar contamination
\citep{Webster1995, White2003}. Based on their findings,  \citet{Webster1995}
and \citet{White2003} suggested that up to $\sim$80\% of the total quasar
population could be made up of red quasars that were missed by previous optical selection
methods. To investigate whether large numbers of dust obscured quasars are
indeed missed, \citet{Glikman2007} combined the VLA Faint Images of the Radio
Sky at Twenty centimeters (FIRST; \citeauthor{Becker1995} 1995) radio survey
and the 2 Micron All Sky Survey (2MASS; \citeauthor{Skrutskie2006} 2006) in an
attempt to counteract the dust bias. However, this introduced a bias toward
the radio-bright red quasar population, as pointed out, for example, by
\citet{Heintz2016} and \citet{Glikman2018}. Combining the Wide-Field Infrared
Space Explorer (WISE; \citeauthor{Wright2010} 2010) with the Sloan Digital Sky
Survey (SDSS; \citeauthor{York2000} 2000) and its extension from the Baryon
Oscillation Spectroscopic Survey (BOSS; \citeauthor{Dawson2013} 2013),
\citet{Glikman2022} found it more likely that the red quasars comprise
$\sim$40\% of the total quasar population. 

Using the SDSS DR7 Quasar catalog \citep{Shen2011}, \citet{Klindt2019}
investigated the fundamental differences between red and blue quasars seen in
radio wavelengths. Searching for mid-IR (MIR) counterparts and checking for
luminosity and redshift effects, they cross-matched the SDSS DR7 Quasar catalog
with sources in the FIRST survey and found a fraction of the
radio detection among the red quasars that was about a factor of 3 higher than that of the blue quasars. \citet{Klindt2019}
claimed that this provided evidence against the orientation-dependent model
because radio emission is not affected by dust, and hence, orientation alone cannot explain the observed differences. In a follow-up study, \citet{Fawcett2020}
found that the radio excess decreased toward the radio-quiet part of the red
quasar population, but concluded that the excess is still significant.
Furthermore, \citet{Fawcett2023} found an intrinsic relation between dust
reddening and the production of radio emission in quasars when studying MIR and
optical-color selected quasars in the Dark Energy Spectroscopic Instrument
(DESI) survey. This relation is such that low-powered jets, winds, or outflows
are thought to cause shocks in the dusty quasar environment, which then powers
radio emission. However, clues pointing toward a lack of radio detection
excess have also been presented previously (see, e.g., \citet{Krogager2016}). When
\citet{Krogager2016} compared their red quasar radio detection fraction to the
radio detection fraction of the red quasars overlapping in SDSS+BOSS, the
overlapping quasars showed a significantly higher radio detection fraction, which
demonstrates that the radio excess might be a selection artifact.

In this paper, we examine the entire sample of quasars found by
\citet{Fynbo2013}, \citet{Krogager2015,Krogager2016}, and \citet{Heintz2020}.
This sample contains quasars that have been missed by classic quasar selection
methods, such as those used to build the SDSS DR7 Quasar catalog.
\citet{Fynbo2013} showed that these quasars differ from the classically
selected quasars by an increased dust reddening. Furthermore, the dust-reddening
was not found to be caused by intervening absorbers, but was instead shown to
be primarily a consequence of the quasar host galaxy. In
Section~\ref{sec:sample}, we outline how we gathered the sample, and in
Section~\ref{sec:method}, we describe how we ensured that color was the only
observed difference between our red quasar sample and the blue SDSS DR7 quasars
with which we compared our sample. We present the results of our investigation in
Section~\ref{sec:results}. Throughout Section~\ref{sec:results}, we also
compare our results to the blue SDSS DR7 quasars in order to search for fundamental
differences between red and blue quasars. The comparison of our results to the
orientation and evolution model is presented in Section~\ref{sec:disc}, where
we also discuss the other parameters that might influence our findings. Following
the \citet{Planck2020}, we assume a flat $\Lambda$CDM cosmology with $H_0=67.4$
km~s$^{-1}$~Mpc$^{-1}$, $\Omega_\Lambda=0.685$, and $\Omega_M=0.315$ throughout
the paper. We use Vega magnitudes throughout the paper, except for
Section~\ref{sec:lumdep}, where we convert into AB magnitudes to calculate the
rest-frame 6 $\mu$m luminosity.

\section{Sample}
\label{sec:sample}
We built our sample of quasars from the catalog of quasar candidates provided in \citet[hereafter F13]{Fynbo2013}, \citet[hereafter K15]{Krogager2015}, \citet[hereafter K16]{Krogager2016}, and \citet[hereafter H20]{Heintz2020}.  
For each candidate, we have broadband photometry from the $u$ (3543 Å), $g$ (4770 Å), $r$ (6231 Å), $i$ (7625 Å), and $z$ (9134 Å) bands from SDSS and the $J$, $H$, $K$ bands from UKIRT Infrared Deep Sky Survey (UKIDSS; \citeauthor{Lawrence2007} 2007). UKIDSS uses the Wide Field Camera (WFCAM) on the 3.8 m United Kingdom Infra-red Telescope (UKIRT). In some cases ($\sim10\%$ of the objects), the NIR band was instead taken from the VISTA \citep{Emerson2006, Dalton2006} Kilo-degree Infrared Galaxy (VIKING; \citeauthor{Edge2013} 2013) survey. The candidates from F13 all have $0.8<g^*-r^*<1.5$ and $r^*-i^*>0.2$, while the candidates from K15 and K16 have $0.5<g^*-r^*<1.0$ and $g^*-r^*>0.5$, respectively. H20 have relied purely on an astrometric selection of quasars as stationary sources in the Gaia survey.  

The sample also contains quasars that have not been published previously, and a few that have been published in single-object studies \citep{Fynbo2017,Heintz2018,geier2019,Fynbo2020}. These quasars were selected with a combination of optical colors as in K16 and astrometric exclusion of stellar sources as in H20. We refer to \citet{Heintz2018} and \citet{geier2019} for further details on the selection of these quasars. The unpublished parts of the survey make up $\sim$50\% of the total survey. 

None of the papers claim to provide an unbiased selection method. On the contrary, they specifically searched for reddened quasars with the main motivation of finding foreground dusty damped Lyman-$\alpha$ absorbers (DLAs). Radio emission has not been part of the selection criteria for any of the quasar candidates. It should also be noted that the quasars were selected specifically to not appear in SDSS spectroscopic database. In total, 550 of the 578 quasar candidates turned out to be quasars after follow-up spectroscopy.

In order for the quasars to be included in our sample, we also introduced the requirement of a spectroscopic redshift. These redshifts are obtained as part of dedicated spectroscopic observations with visually inspected redshift determinations for all objects. We refer to the list of references for the surveys in the appendix for further details. Five hundred and forty-three of the total 550 quasars fulfill this requirement. In addition to the broadband photometric data and spectroscopic redshifts, we also searched for MIR counterparts of the quasars using WISE \citep{Wright2010} and radio counterparts using FIRST \citep{Becker1995, Becker2012, Helfand2015}. FIRST operates at a frequency of 1.4 GHz (20 cm wavelength) and offers a high sensitivity of typically about 1 mJy for point sources. With an angular resolution as fine as 5 arcsec, FIRST covers over 10,000 deg$^2$ of the northern sky in a region that is largely coincident with SDSS. For the detections, we adopted a 10 arcsec cross-matching radius. We took the closest-distance match within the 10 arcsec to be the radio counterpart. It should be noted that the largest match distance we found is 1.2 arcsec. In the case of multiple sources within the 10 arcsec search radius, we also checked SDSS for the existence of multiple optical sources. This was only the case for one of the quasars: CQ0155+0438. Hence, we did not include CQ0155+0438 in the sample. We find that 534 of the redshift-confirmed quasars have MIR counterparts, while a small subset of 33 quasars have radio counterparts detected in FIRST. We consider the 534 MIR counterpart quasars to be the parent sample for our study here and call this sample the eHAQ+GAIA23 sample. We chose this name because the sample contains sources from the HAQ Survey, the eHAQ Survey, and the Gaia quasar survey. The number 23 represents the year 2023, in which this project was carried out. See the appendix for details on accessing the published data. \\


\section{Methods}
\label{sec:method}
In order to facilitate a direct comparison to the blue SDSS DR7 quasars, we followed the method in \citet{Klindt2019}. Their method consists of two steps taken to reduce the differences that are not related to color between the red and blue SDSS DR7 quasar population. First, they used $g^*-i^*$ color quantile cuts to define a red and a blue quasar population. Then, they used an unsupervised nearest-neighbor algorithm in the rest-frame 6 $\mu$m luminosity-redshift space to match every red quasar with a blue one. These steps are performed to ensure that the populations are similar in all other parameters apart from color, which means that any differences can be attributed to the color of the quasars. Redshift and 6 $\mu$m luminosity are chosen above other dependences because they have a potentially high impact on the quasar color. When we do not take redshift into account, a high-redshift blue quasar might seem red. The 6 $\mu$m rest-frame luminosity is a measure of the thermal emission from warm–hot dust in the quasar torus. It is heated by the accretion disk emission and indicates the central black hole accretion level, which is correlated to the bolometric luminosity of the quasar \citep{Hickox2018}. By matching this luminosity distribution for the two different color populations, we ensure that the quasar activity levels are similar.

\subsection{Redshift dependence}
\label{sec:reddep}
One of the keystones in exploring the differences between red and blue quasars is to ensure that the only observed difference in the sample is the color. To do this, we applied several cuts on the sample redshift. We divided the eHAQ+GAIA23 sample into two redshift bins: $z_1$, including quasars with $0.8<z\leq1.5$, and $z_2$, including those with $1.5<z\leq2.4$. However, we also decided to study $z_3$, including those with $2.4<z\leq4.25$. The resulting number of quasars within each redshift bin is presented in Table~\ref{tab:parent}.  
\begin{table}
    \centering
    \resizebox{\columnwidth}{!}{%
    \begin{tabularx}{\linewidth}{lXXXX}
        \toprule
        \toprule
         Sample \hfill & N$_\text{all}$ \hfill  & N$_{z1}$ \hfill  & N$_{z2}$ \hfill  & N$_{z3}$  \\ \midrule \midrule
         Redshift & [0.09;4.25] & (0.8;1.5] & (1.5;2.4] & (2.4;4.25] \\\\
         eHAQ+GAIA23 & 534  & 118  & 183  & 202  \\
         BAL & 179 & 10 & 84 & 83 \\ \\
         rQSO & 423 & 107  & 179  & 110  \\
         BAL rQSO & 163  & 10  & 84  & 67  \\
         not-red QSO & 111  & 11  & 4  & 92  \\\\
         FrQSO & 33 & 16  & 10  & 3 \\
         BAL FrQSO & 3  & 0 & 1 & 2 \\
         \bottomrule \bottomrule
    \end{tabularx}%
    }
    \vspace{0.1cm}
    \caption{Number of quasars in the eHAQ+GAIA23 catalog. Column 2 presents the total sample, and columns 3, 4, and 5 present different redshift ranges. We define red quasars (rQSO) as those with a $g^*-i^*$ color value in the upper 10th percentile of the SDSS DR7 Quasar Catalog $g^*-i^*$ color values, while all BAL quasars were inspected visually. All quasars between the upper and lower 10 percentiles are defined as not-red quasars.}
    \vspace{-0.5cm}
    \label{tab:parent}
\end{table}
\subsection{Color dependence}
\label{sec:coldep}
\begin{figure}
\vspace{-0.25cm}
\centering
\includegraphics[width=\hsize, trim={2cm 0cm 0 1.1cm},clip]{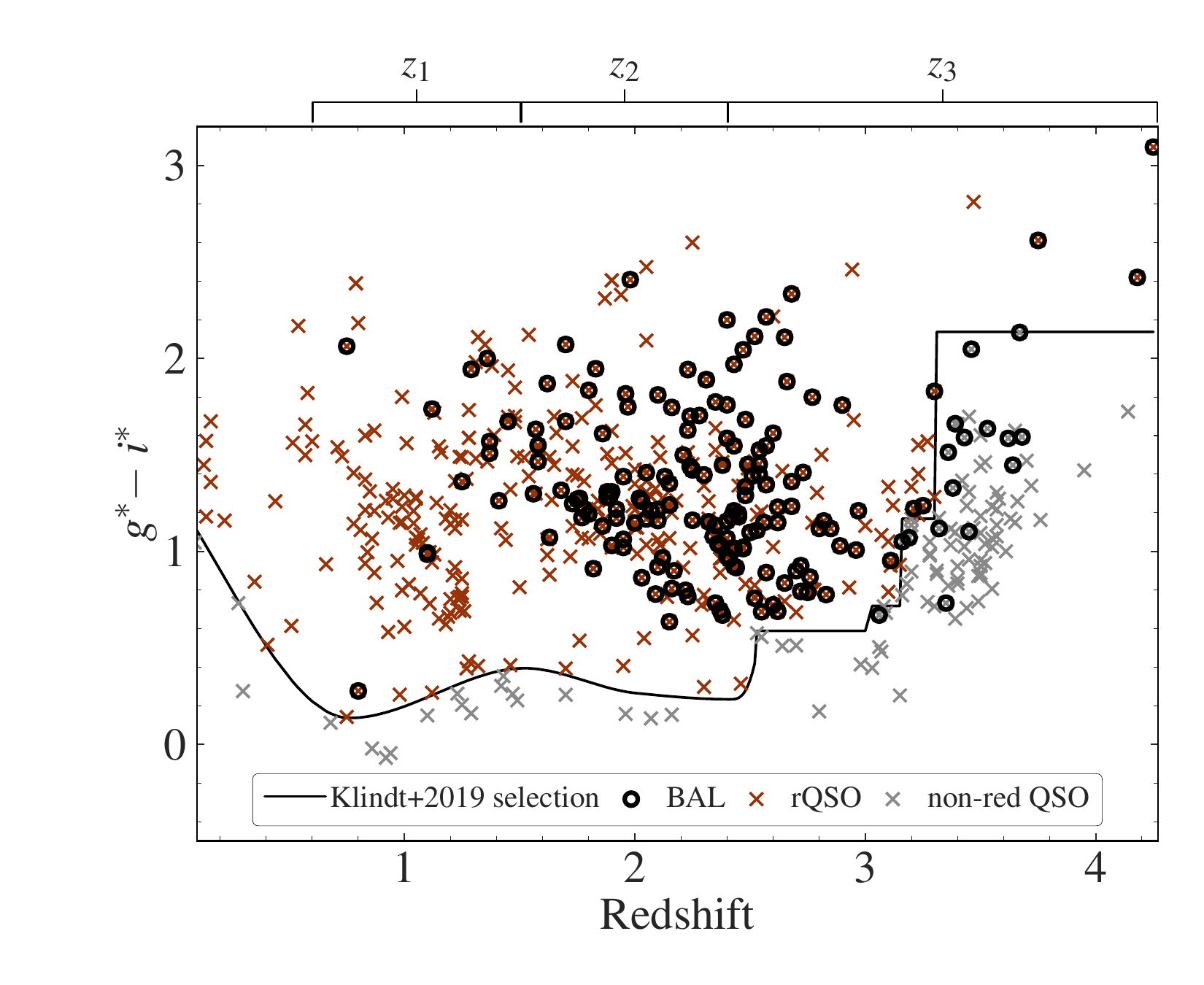}
\caption{Distribution of the $g^*-i^*$ color as a function of redshift for
the quasars in the new quasar catalog. Every cross represents a quasar. The
black line shows the red quasar selection criteria from \citet{Klindt2019},
such that quasars with a higher color value are labeled rQSO and are shown
in red in this figure, while those with a lower color value are labeled
not-red QSO and are shown in gray in this figure. The categorization of the
BAL quasars, marked with black circles, relied on visual assessment without
    the application of rigid standards (see F13, K15, K16, H20).}
\label{fig:redcut}
\end{figure}

We also ensured that we defined a red and a blue quasar in the same manner as
\citet{Klindt2019}. They defined a red quasar as a quasar with a $g^*-i^*$
color value in the upper 10th percentile of the SDSS DR7 quasar catalog
$g^*-i^*$ color values, while a blue quasar should belong to the lower 10th
percentile of the SDSS DR7 QC $g^*-i^*$ color values. All quasars in between
these percentiles are defined as not-red control quasars. \citet{Klindt2019}
already noted that this might lead to issues in the $z_3$ bin, where the $g^*$
band is affected by the Lyman-$\alpha$ break. For this reason, we conducted our
analysis with and without the $z_3$ bin. In Figure~\ref{fig:redcut}, we show
the color selection as a black line on top of the $g^*-i^*$ color versus
redshift distribution of each quasar in the eHAQ+GAIA23 sample.  
The number of red quasars (rQSO) and red quasars with a radio detection (FrQSO)
can be found in Table~\ref{tab:parent}. It is evident from this table that 423
of the eHAQ+GAIA23 quasars are red, and the remaining 111 are not red. Upon
inspecting the distribution of these two quasar classes across the three
redshift bins, we observe that the red quasars are predominantly concentrated
in the middle redshift bin, where $\sim$40\% of them are found. The not-red
quasars in eHAQ+GAIA23 are primarily located in the highest redshift bin, where
$\sim$80\% of them are found. One explanation for this could be that SDSS
contains a smaller fraction of red quasars at higher redshifts, and hence the
quantile cut is pushed toward higher $g^*-i^*$ values.

As can be seen from Table~\ref{tab:parent}, 179 of the eHAQ+GAIA23
quasars qualify as BAL, meaning that this new catalgo contains $\sim34\%$ BAL
quasars. For the red quasars alone, we find 163 red BAL quasars out of 423 red
quasars. For the red quasars, the BAL percentage therefore increases to
$\sim39\%$ BAL quasars. For comparison, we include the BAL number statistics
from the SDSS DR7 quasar catalog \citep{Shen2011} in Table~\ref{tab:balstat}.
\begin{table}
    \centering
    \begin{tabularx}{\linewidth}{lYY}
        \toprule
        \toprule
         Sample  & Num. BAL  & \multicolumn{1}{c}{BAL Percentage (\%)}    \\
         \midrule \midrule
         Parent & 179 & 34 \\
         rQSO & 163 & 39  \\
         rQSO $\in z_1 \cup z_2$  & 94 & 33  \\
         SDSS DR7 & 6214 & 5.9 \\ \bottomrule \bottomrule
    \end{tabularx}
    \vspace{0.1cm}
    \caption{Statistics of the quasar distribution with respect to the number of BALs. For comparison, we include information from the SDSS DR7 quasar catalog \citep{Shen2011}.}
    \label{tab:balstat}
\end{table}
\subsection{Luminosity dependence}
\label{sec:lumdep}

\begin{figure}
     \centering
     \begin{subfigure}[b]{0.4\textwidth}
         \centering
         \includegraphics[width=\textwidth]{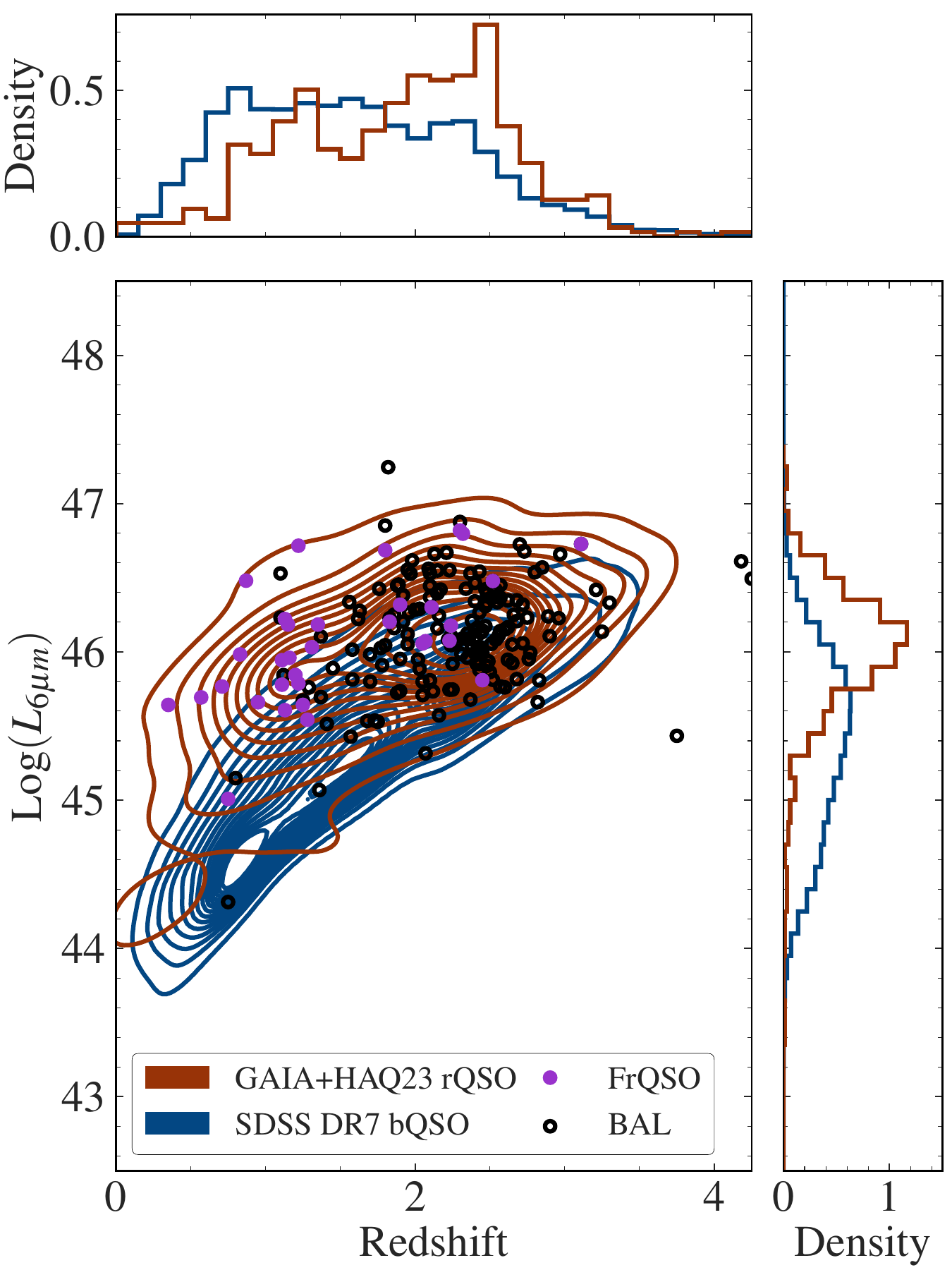}
         \caption{Before the nearest-neighbour matching}
         \label{fig:nonmatchedhist}
     \end{subfigure}
     \hfill
     \begin{subfigure}[b]{0.4\textwidth}
         \centering
         \includegraphics[width=\textwidth]{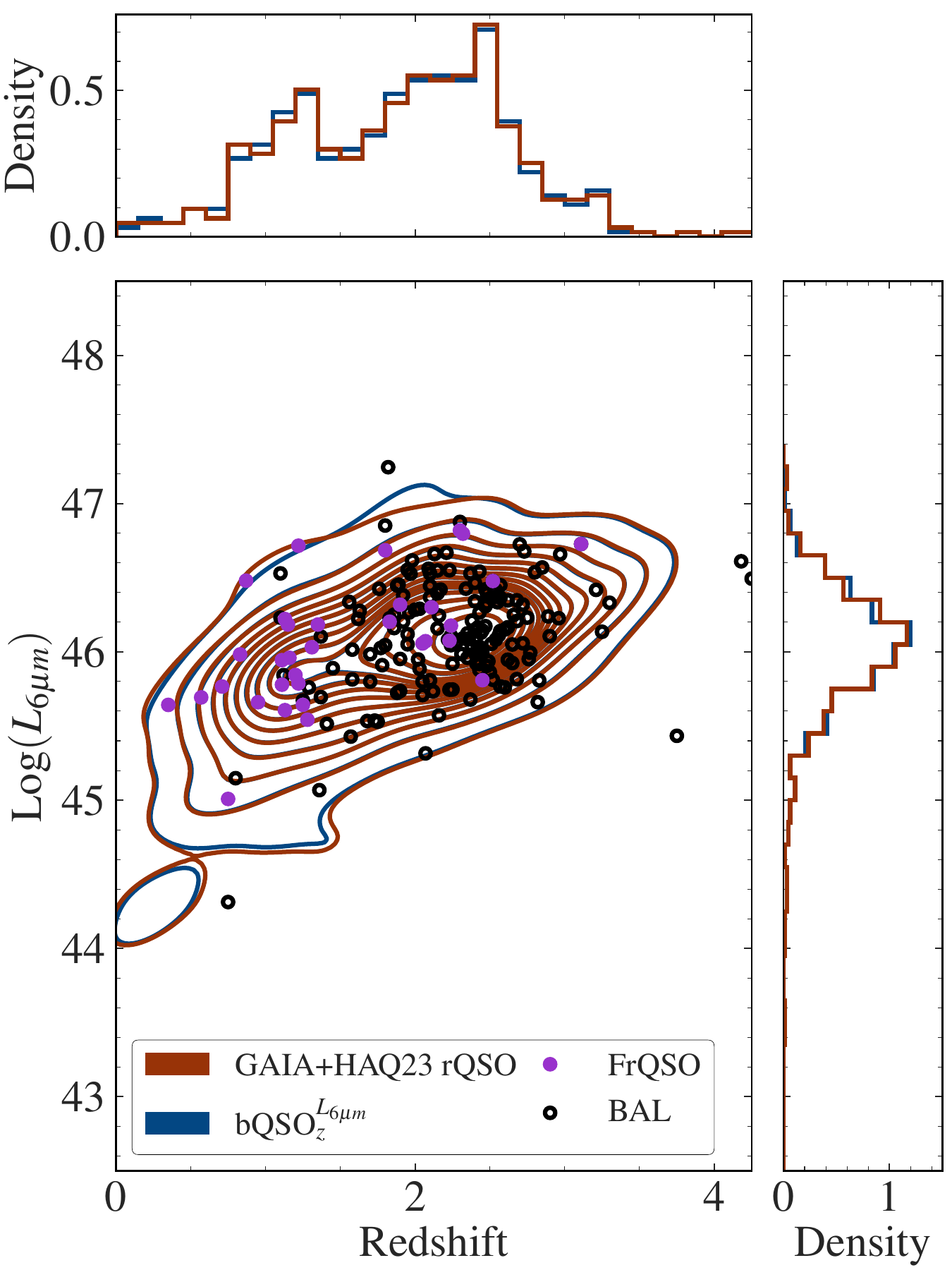}
         \caption{After the nearest-neighbour matching}
         \label{fig:matchedhist}
     \end{subfigure}
     \caption{
The red kernel density estimate plot
represents the red eHAQ+GAIA23 quasar distribution, and each purple dot
represents a radio-detected red quasar. The black circles represent red BAL
quasars. In addition to these markers, the blue kernel density estimate plot
shows the blue SDSS DR7 quasars. 
Center: Logarithm of the rest-frame 6 $\mu$m luminosity of either
all the blue SDSS quasars (a) or the redshift-luminosity matched blue SDSS
quasars (b) in blue curves and the red eHAQ+GAIA23 quasars in red curves
as a function of redshift, assuming a continuous probability density
curve. Top: Density histogram of the redshift for the two different
populations. Right: Density histogram of the rest-frame 6 $\mu$m
luminosity for the two different populations.}
\label{fig:double}
\end{figure}

To ensure that any difference between the red and blue quasars cannot be
explained by differences in the rest-frame luminosities, we determined the
rest-frame 6 $\mu$m luminosity of the red quasars in our sample and compared
them to the rest-frame 6 $\mu$m luminosities of the red and blue SDSS quasars.
We used the MIR-magnitude data from WISE. First, we converted from the WISE
Vega-system into AB magnitudes using $m_\text{AB} = m_\text{Vega} + \Delta m$,
where $\Delta m=3.339$ for the W2 band and $\Delta m=5.174$ for the W3 band
\citep{Tokunaga2005}. Under the assumption that the magnitudes are
monochromatic at the effective filter wavelength, we converted from AB
magnitudes into flux densities following the definition given by
\citet{Tokunaga2005}. We performed a log-linear fit between the W2 and W3
effective filter wavelength flux densities, which we used to either interpolate
or extrapolate the rest-frame 6 $\mu$m flux density, depending on the quasar
redshift. It should be noted that for the highest-redshift quasar, we
extrapolated up to $\sim$ 30 $\mu$m. At this wavelength, the extrapolation is
not necessarily reliable. The rest-frame 6 $\mu$m luminosity $L_{6\mu
\text{m}}$ distribution as a function of redshift is shown in
Figure~\ref{fig:nonmatchedhist}. 

With respect to ensuring that the only
observed difference between our red quasars and the SDSS blue quasars is the
color, we need to correct this difference in the luminosity distributions.
Without any form of correction, the difference prohibits us from excluding that
any other difference observed between the two populations is more than the mere
result of a correlation with the rest-frame 6 $\mu$m luminosity. 

Even though the redshift distributions of the blue SDSS population and the red
eHAQ+GAIA23 population look similar, the top histogram in
Figure~\ref{fig:nonmatchedhist} shows a shift toward higher redshifts for our
red quasars. The right histogram in Figure~\ref{fig:nonmatchedhist} reveals a
clear difference in the two luminosity distributions. While the red eHAQ+GAIA23
quasar sample has a luminosity median of $\log(\overline{L}_{6\mu
\text{m}})=46.0^{+0.3}_{-0.4}$, the blue SDSS population has a lower and
visibly broader luminosity distribution with median $\log(\overline{L}_{6\mu
\text{m}})=45.4^{+0.6}_{-0.7}$. In order to remove this difference and be able
to report differences related solely to quasar color, we followed an approach
similar to the one taken by \citet{Klindt2019}: We ran the red eHAQ+GAIA23
quasars through a \texttt{scikit-learn} \citep{scikit-learn}, unsupervised
nearest-neighbor algorithm in the rest-frame 6 $\mu$m luminosity-redshift
space. For each of our red sample quasars, the algorithm found an SDSS blue
quasar within a fixed tolerance of 0.1 dex in luminosity space and 0.1 in
redshift space. After using the nearest-neighbor algorithm, the
redshift-luminosity matched blue quasars show a rest-frame 6 $\mu$m luminosity
distribution with median $\log(\overline{L}_{6\mu
\text{m}})=46.0^{+0.3}_{-0.4}$.

Figure~\ref{fig:matchedhist} shows the kernel density estimate plot after using
the nearest-neighbor algorithm. This served mainly as a sanity check and
revealed that the approach has the intended effect. The redshift-luminosity
matched blue quasars show a rest-frame 6 $\mu$m luminosity distribution with
median $\log(\overline{L}_{6\mu \text{m}})=46.0^{+0.3}_{-0.4}$.

\begin{figure}
    \centering
    \includegraphics[width=1.1\columnwidth]{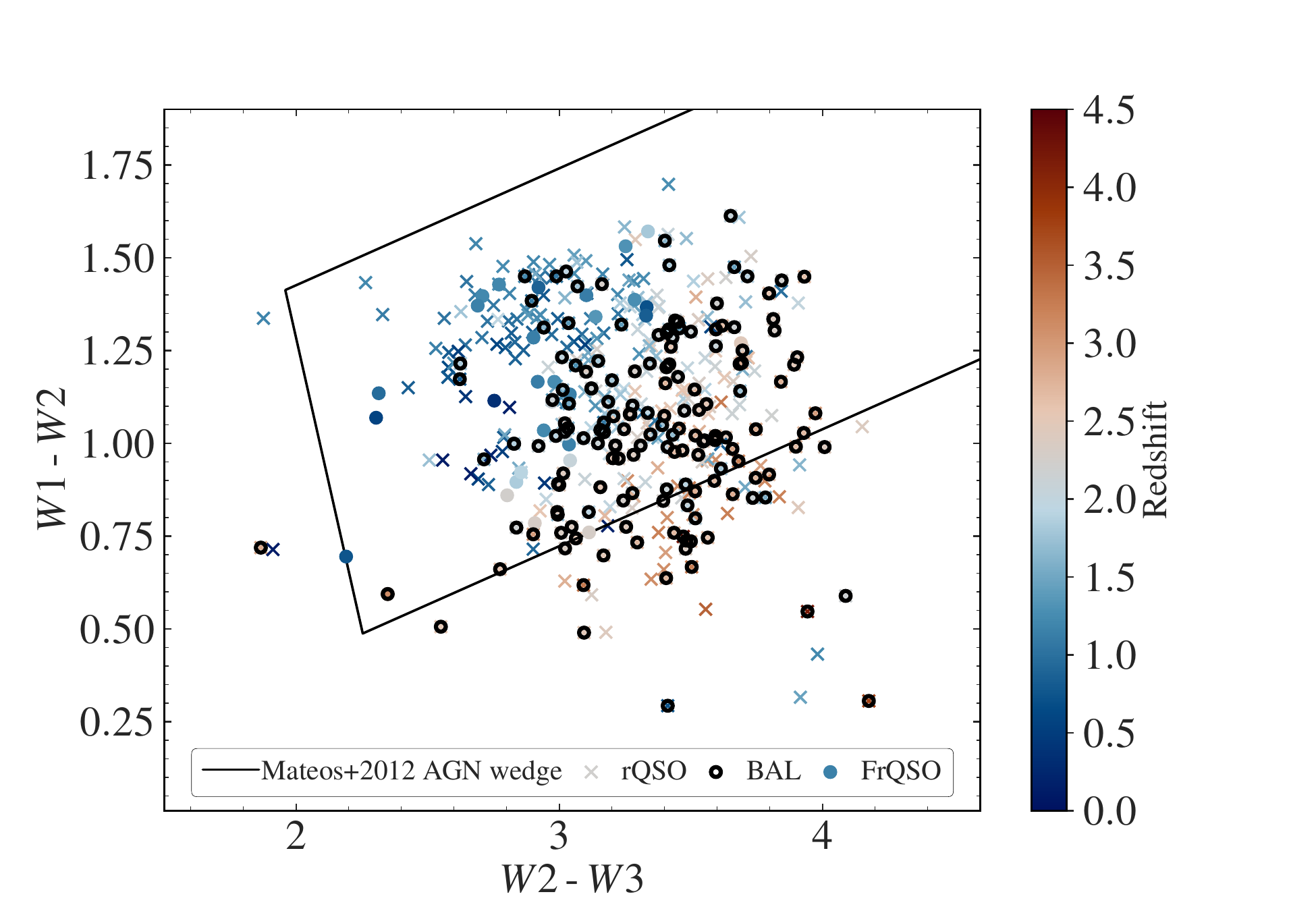}
    \caption{Red quasar distribution in the WISE $(W2-W3$,$W1-W2)$ color space. Each cross is a red quasar, each dot is a radio quasar, and each black circle indicates whether the red quasar is a BAL quasar. The color of the markers depends on the quasar redshifts.
              }
         \label{fig:wedge}
\end{figure}

\subsection{Host galaxy dependence}
\label{sec:agnwedge}
In Figure~\ref{fig:wedge}, the distribution of the red quasars in the
eHAQ+GAIA23 sample is shown in the $W2-W3$ vs. $W1-W2$ color space alongside
the AGN wedge from \citet{Mateos2012}.  Each cross represents a red quasar,
while the color of the cross is determined by the quasar redshift. The dots
show the distribution of the red quasars with radio detection, and a black
circle marks which of the red quasars are BAL quasars. Table~\ref{tab:agnwedge}
shows that 59 of the 423 red quasars lie outside the AGN wedge. The highest
percentage is found in the $z>2.4$ sample. Even though the AGN torus is
expected to dominate the emission at MIR wavelengths, emission from star
formation in the host galaxy might still contaminate the spectrum. If the MIR
fluxes are indeed contaminated by the host galaxy, this would bias our
rest-frame 6 $\mu$m luminosity calculations. The AGN wedge systematically
deselects spectra for which the host galaxy star-formation contributes more
than 10\% to the MIR luminosity emitted by the AGN. Spectra with a
contamination from star formation luminosity $>10\%$ end on the lower right
side of the wedge \citep{Mateos2012}. This explains that the highest percentage
of quasars outside the AGN wedge is found in the highest-redshift sample, which
is a more active part of the star formation history \citep{Madau2014}. The red
BAL quasars show a similar pattern, with an outside percentage peaking at
$z_3$. Only the small subcategory of red quasars with a radio detection have no
outliers with respect to the AGN wedge.

\begin{table}
    \small
    \centering
    \resizebox{1.0\columnwidth}{!}{%
    \begin{tabularx}{\linewidth}{lcccc}
        \toprule
        \toprule
         Sample \hfill & Num. outside \hfill  & \multicolumn{3}{c}{Percentage outside (\%), $z$ $\in$ \hfill}    \\
           &    & $(0.8;1.5]$ & $(1.5;2.4]$   & $(2.4;4.25]$  \\
         \midrule \midrule
         rQSO & 59 & 3.7  & 6.7  & 36\\
         bQSO$^{L6\mu m}_z$ & 74 & 1.8 & 6.8 & 51 \\
         BAL rQSO & 32 & 0 & 10 & 34 \\
         FrQSO & 0 & 0  & 0 & 0 \\
         FbQSO$^{L6\mu m}_z$ & 4 & 0 & 0 & 66 \\ \bottomrule \bottomrule
    \end{tabularx}%
    }
    \vspace{0.1cm}
    \caption{Statistics of the red and redshift-luminosity matched blue quasar distribution with respect to the AGN wedge in WISE colors as defined by \citet{Mateos2012}. Information of the subcategories of red BAL quasars (BAL rQSO), radio-detected red quasars (FrQSO), and the FIRST-detected percentage of the redshift-luminosity matched blue SDSS quasar population (FbQSO$^{L6\mu m}_z$) is also included.}
    \label{tab:agnwedge}
\end{table}

The statistics presented in Table~\ref{tab:agnwedge} generally have a higher
number of quasars outside the AGN wedge compared to the SDSS population of blue
quasars, but not compared to the redshift-luminosity matched subpopulation. The
redshift-luminosity matched subpopulation has a higher percentage of quasars
outside the AGN wedge at redshifts $z>1.5$. The comparison between red and blue
quasars was to be as unaffected by other physical parameters as possible.
Therefore, we proceeded to run the unsupervised nearest-neighbor algorithm on
the red quasars within the AGN wedge and drew neighbors from the full blue SDSS
quasar population. This yielded a rest-frame 6 $\mu$m luminosity distribution
with median $\log(\overline{L}_{6\mu \text{m}})=46.0^{+0.3}_{-0.4}$. This is
similar to what we found when we included quasars outside the AGN wedge. 


\section{Results: Radio-detection rates}
\label{sec:results}
In Table~\ref{tab:agnfrac}, we report the FIRST-detected percentage in each of the three redshifts bins for the two different subsamples of red eHAQ+GAIA23 quasars described in Section~\ref{sec:coldep} and~\ref{sec:agnwedge}. We also report the FIRST-detected percentage for the luminosity-redshift matched SDSS sample of blue quasars described in Section~\ref{sec:lumdep}. All values were calculated using the Bayesian binomial confidence interval (CI) method described in \citet{Cameron2011}, such that the reported value is the 50\% CI and the lower and upper uncertainty are the 15.87\% and 84.13\% CI ($\pm$ 1$\sigma$), respectively. The leftmost column displays the FIRST-detected percentage of the red quasars (rQSO) selected as described in Section~\ref{sec:coldep}, while the center column displays the FIRST-detected percentage of the red quasars within the AGN wedge (AGN-w rQSO). As shown in Figure~\ref{fig:nonmatchedhist}, these percentages should not be compared directly to the blue quasar FIRST-detected percentages reported by \citet{Klindt2019} because of the differences in the rest-frame 6 $\mu$m luminosity distributions. Instead, we report the FIRST-detected percentage of the redshift-luminosity matched blue quasar population in the rightmost column of Table~\ref{tab:agnfrac}.

\begin{table}
    \centering
    \begin{tabularx}{\linewidth}{lYYY}
        \toprule
        \toprule
         \multicolumn{1}{c}{Redshift}  & \multicolumn{3}{c}{FIRST-detected percentage (\%)}     \\
           &  rQSO  & \multicolumn{1}{c}{AGN-w rQSO} & \multicolumn{1}{c}{bQSO$^{L6\mu m}_z$}     \\
         \midrule \midrule
         \multicolumn{1}{c}{$0.8<z\leq$1.5} & \multicolumn{1}{c}{15.3$^{+3.7}_{-3.2}$}  & \multicolumn{1}{c}{16.0$^{+3.8}_{-3.4}$} & 13.2$^{+3.4}_{-3.0}$  \\
         \multicolumn{1}{c}{$1.5<z\leq2.4$} & \multicolumn{1}{c}{5.9$^{+1.9}_{-1.6}$} & \multicolumn{1}{c}{6.3$^{+2.0}_{-1.7}$} & 6.0$^{+1.9}_{-1.6}$ \\ 
         \multicolumn{1}{c}{$z>2.4$} & \multicolumn{1}{c}{2.9$^{+1.7}_{-1.2}$} & \multicolumn{1}{c}{5.1$^{+3.0}_{-2.2}$} & 5.8$^{+2.4}_{-1.9}$ \\
         \bottomrule \bottomrule
    \end{tabularx}
    \vspace{0.1cm}
    \caption{FIRST-detected percentage in each of the redshift bins for two subsamples of the new catalog: The red quasars (rQSO), and the red quasars within the AGN wedge (AGN-w rQSO). The rightmost column shows the FIRST-detected percentage of the redshift-luminosity matched blue SDSS quasar population.}
    \label{tab:agnfrac}
\end{table}

\begin{figure}
    \centering
    \includegraphics[width=\hsize]{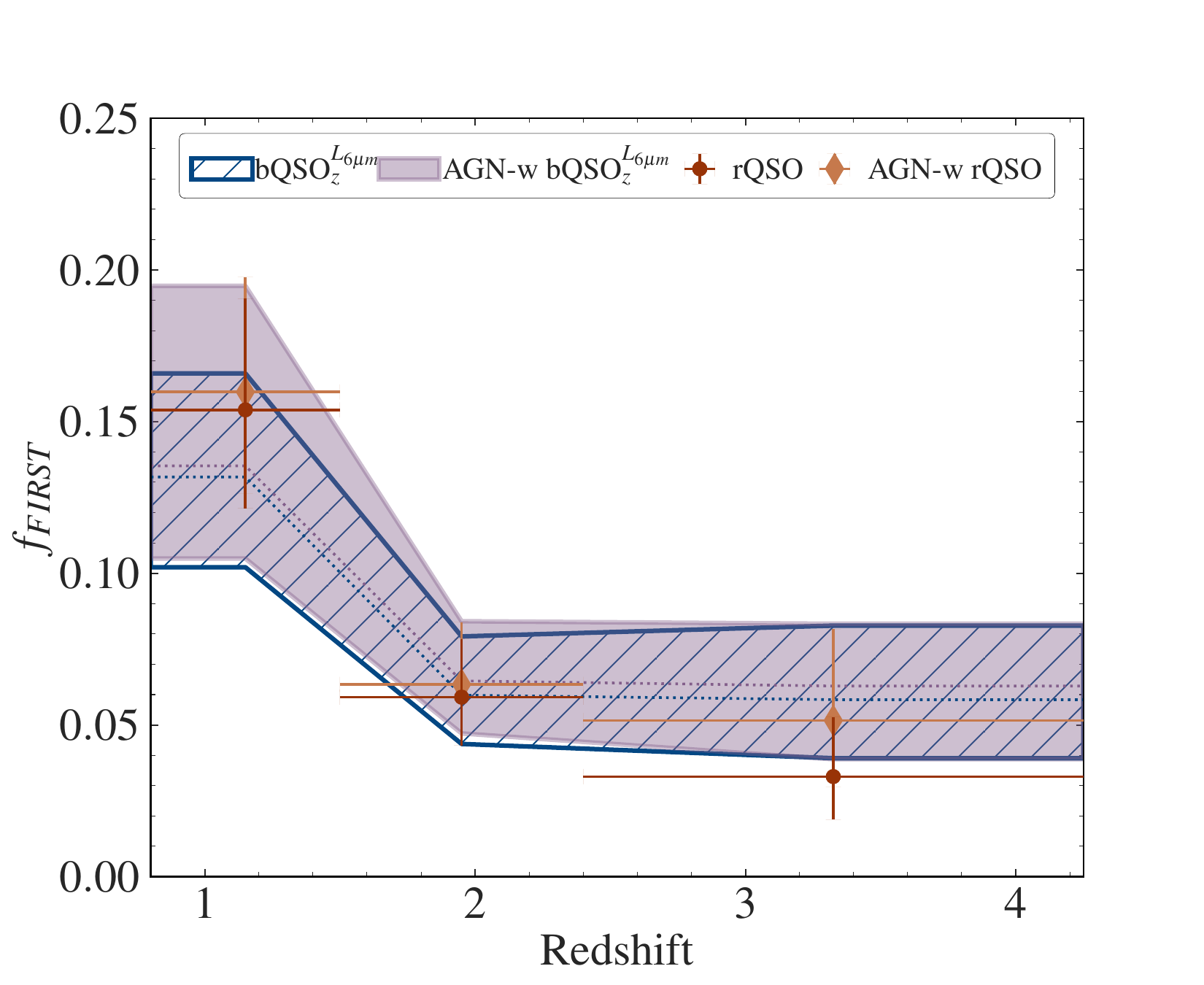}
    \caption{FIRST-detected fraction as a function of redshift for the two different subsamples of our new catalog: The red quasars (rQSO) shown in dark red, and the red quasars within the AGN wedge (AGN-w rQSO) shown in orange. The luminosity-redshift matched blue quasars from SDSS are also included. The area highlighted by blue diagonal stripes corresponds to their detection fraction $\pm1\sigma$ confidence intervals. The shaded purple area highlights the area corresponding to the blue SDSS quasars matched with redshift-luminosity to our red quasars within the AGN wedge. The two dashed lines show the median of the luminosity-redshift matched blue quasars from SDSS and the blue SDSS quasars matched with redshift-luminosity to our red quasars within the AGN wedge.
              }
         \label{fig:firstredshift}
\end{figure}

The values reported in Table~\ref{tab:agnfrac} are plotted in Figure~\ref{fig:firstredshift}. In this figure, the dark red dots with error bars represent the fraction of FIRST-detected red quasars and their 1$\sigma$ upper and lower limits. The orange diamonds and error bars represent the fraction of FIRST-detected red quasars with observed colors within the AGN wedge and their limits. The area shown by the blue hashed region highlight the area corresponding to the fraction of FIRST-detected redshift-luminosity matched blue SDSS quasars $\pm1\sigma$ errors. Similarly, the shaded purple region highlight the area corresponding to the fraction of FIRST-detected blue SDSS quasars matched with redshift-luminosity to our red quasars within the AGN wedge.

Figure~\ref{fig:firstredshift} illustrates that the red quasars in the eHAQ+GAIA23 sample do not show a higher radio detection rate within $1\sigma$ than the blue quasar sample for any redshifts. Even when we exclude the red quasars outside the AGN wedge, the detection fraction of radio emission is the same within the $1\sigma$ confidence intervals compared to the blue quasar sample in all redshift bins. Even though it is not statistically significant, we note that the red quasar values are at the top part of the blue quasar distribution in the lowest-redshift bin. In the highest-redshift bin, the red quasar values lie in the lower part of the blue quasar distribution, even though this is not statistically significant either.


\section{Discussion and conclusions}
\label{sec:disc}
We have presented an analysis of the radio-detection fraction of the red quasars in eHAQ+GAIA23. Using a color-cut criterion in the $g^*-i^*$ color space, we selected only quasars that would belong to the upper 10th $g^*-i^*$ color percentile of quasars in the SDSS quasar catalog. Neither of the redshift bins show an excess of sources with radio emission detected in FIRST compared to redshift-luminosity matched SDSS blue quasars. 

\subsection{Quasar redness models}
\label{sec:redmodels}
Orientation-dependent models claim that the redness of quasars can be explained by the viewing angle with respect to the dusty torus \citep{Antonucci1993,Urry1995,Netzer2015}. In this framework, red quasars are simply observed with an inclination angle and hence through the dusty torus, while blue quasars are observed face-on. While the intervening dusty torus will change the optical colors, the radio emission remains unaffected. Under the assumption that the dusty torus is the only cause of the observed redness, there should thus be no difference in the radio emission from red and blue quasars. In this manner, our findings support the orientation model for quasar redness.

On the other hand, the evolutionary models attribute the redness of quasars to different phases in the life of quasars. A red quasar evolves from an early-phase blue quasar that has developed strong winds in order to drive its initial layers of dust and gas away and reveal the hidden accretion disk within. Thus, these models suggest that red quasars have stronger winds than blue quasars. Strong winds have been associated with weak radio emission \citep{Mehdipour2019}, which suggests that red quasars should have fewer detected sources at a constant detection sensitivity. This is not what we observe for our sample, and hence our findings only provide evidence for the orientation-dependent model for quasar redness.

\subsection{The BAL Effect}
\label{sec:bal}
The strong winds of the evolutionary model framework are a well-known characteristic of BAL quasars. Following the results of \citet{Mehdipour2019}, it is expected that BAL quasars are more frequently radio quiet than other quasars. In the context of explaining why the red eHAQ+GAIA23 quasars have a similar radio detection fraction compared to their luminosity-redshift matched blue SDSS quasars, we recall from Table~\ref{tab:balstat} that our red quasar sample has a high BAL fraction even at redshifts $z<2.4$. With a higher fraction of radio-quiet quasars, it is expected that the radio-detection fraction decreases. In other words, it is possible that the radio-detection fraction of the red eHAQ+GAIA23 quasars is lower solely because of the higher BAL quasar fraction. This links the question of whether the radio detection fraction depends on quasar color to the question of the intrinsic BAL quasar fraction in the total quasar population. The BAL quasar fraction remains an open question, but the recent study by \citet{Bischetti2023} estimated it to increase with redshift from $\sim20\%$ at redshift $z\sim2-4$ to $\sim50\%$ at redshift $z\sim6$. An exploration of the BAL fraction of the total red quasar population would require a selection cut that is unbiased against BAL and red quasars.

\subsection{Selection effects}
\label{sec:select}
The quasars in eHAQ+GAIA23 are not selected to be unbiased. Instead, the selection criteria aim to target red quasars, in particular, to explore how frequently quasars are reddened by dusty foreground absorbers at redshifts $z\gtrsim2$ \citep{Krogager16b,geier2019}. Hence, the chosen selection cuts are also expected to play a role in the results presented in this work. The HAQ Survey presented by F13 led the later surveys and particularly the eHAQ Survey to an increased focus on targeting fewer quasars at redshifts $z<2$ with high dust reddening as well as fewer unobscured quasars at redshifts $z>3.5$. This resulted in a decrease of intrinsically red quasars at lower redshifts, especially at redshifts $z\lesssim1.5$. One of the main reasons for doing so is to ensure the ability to detect Ly-$\alpha$ in absorption in the observed spectra from the surveys, which had posed a problem for the HAQ Survey (see K16).

Even though the original purpose of the survey was aided by this, the intended bias subtly compromises our results for red quasars with redshifts $0.8<z\leq1.5$. In addition to explaining why the combined surveys have fewer quasars in this redshift bin, the focus might also explain why this population behaves more similar to the red SDSS quasars with radio-detection fractions in the upper part of the luminosity-redshift matched blue SDSS quasar distribution. 

The DESI survey quasars studied, for example, by \citet{Fawcett2023} also behave more similar to the SDSS quasars. These quasars are selected based on MIR and optical color and display an intrinsic relation between dust reddening and the production of radio emission. In contrast to the eHAQ+GAIA23 quasars, the DESI survey quasars might not be selected completely independently from the SDSS quasars. In addition to color cuts in the MIR and optical colors, \citet{Chaussidon2023} used a random forest machine-learning algorithm to improve the success rate for DESI with respect to quasars. This algorithm is trained on quasars known from the SDSS Stripe 82, which opens the possibility for a correlation between the SDSS and the DESI quasars. Hence, the need for a color-unbiased quasar sample remains.

\subsection{Conclusion and future work}
Our analysis of the radio-detection fraction of red quasars in the eHAQ+GAIA23 sample, when compared to the SDSS blue quasars, provides significant insights into the nature of quasar redness. Our findings predominantly support the orientation-dependent model, suggesting that the redness of quasars is primarily a result of the viewing angle relative to the dusty torus. This conclusion is drawn from the observation that there is no significant difference in the radio emission between red and blue quasars. This aligns with the expectation that the dusty torus affects optical colors without influencing radio emission. In contrast, the evolutionary model, which links quasar redness to different developmental phases and associates strong winds with weak radio emission, finds less support in our data. The lack of a statistically discernible decrease in radio detection fraction in red quasars, as predicted by the evolutionary model, suggests that the red quasar redness is less likely to be due to this phase-based evolution. Furthermore, the intricacies of our sample selection in eHAQ+GAIA23, with a particular focus on red quasars and the exclusion of certain quasar populations due to survey biases, have implications for our results. While these biases were essential for the previous survey's objectives, they also limit the generalizability of our conclusions, particularly in the context of the broader quasar population.

In future works, we will analyze the red quasar spectra further to provide more details on the emission and absorption lines of this quasar population. On larger timescales, a hope for a future disentanglement of the selection biases discussed here has been presented with the 4-meter Multi-Object Spectroscopic Telescope (4MOST) Gaia purely astrometric quasar survey \citep[4G-PAQS,][]{Krogager2023}. With its wide-field, high-multiplex, optical spectroscopic survey facility \citep{4MOST}, 4MOST will provide a unique opportunity to build the first color-unbiased quasar survey based solely on astrometry from Gaia without an assumption on the spectral shape. However, it should be noted that the 4G-PAQS will still be limited in magnitude due to its selection of Gaia-detected sources. The 4G-PAQS community survey will target a total of approximately 300,000 quasar candidates. In addition to studying quasar feedback through BAL outflows at redshifts $0.8<z<4$, 4G-PAQS will also aim to quantify the dust bias in quasar absorption systems at redshifts $2 < z < 3$. This will provide a larger unbiased sample of both red and not-red quasars to determine the origin of the differences between red and blue quasars. \\


\begin{acknowledgements}
The Cosmic Dawn Center (DAWN) is funded by the Danish National Research
Foundation under grant DNRF140. SV and JPUF is supported by the Independent
Research Fund Denmark (DFF–4090-00079) and thanks the Carlsberg Foundation
for support. LC and GM are supported by the Independent Research Fund
Denmark (DFF 2032-00071). KEH acknowledges support from the Carlsberg
Foundation Reintegration Fellowship Grant CF21-0103. This publication makes
use of data products from the Wide-field Infrared Survey Explorer, which is
a joint project of the University of California, Los Angeles, and the Jet
Propulsion Laboratory/California Institute of Technology, funded by the
National Aeronautics and Space Administration. This publication makes use
of data products from the Two Micron All Sky Survey, which is a joint
project of the University of Massachusetts and the Infrared Processing and
Analysis Center/California Institute of Technology, funded by the National
Aeronautics and Space Administration and the National Science Foundation.
Funding for SDSS-III has been provided by the Alfred P. Sloan Foundation,
the Participating Institutions, the National Science Foundation, and the
U.S. Department of Energy Office of Science. The SDSS-III web site is
\url{http://www.sdss3.org/}. SDSS-III is managed by the Astrophysical
Research Consortium for the Participating Institutions of the SDSS-III
Collaboration including the University of Arizona, the Brazilian
Participation Group, Brookhaven National Laboratory, Carnegie Mellon
University, University of Florida, the French Participation Group, the
German Participation Group, Harvard University, the Instituto de
Astrofisica de Canarias, the Michigan State/Notre Dame/JINA Participation
Group, Johns Hopkins University, Lawrence Berkeley National Laboratory, Max
Planck Institute for Astrophysics, Max Planck Institute for
Extraterrestrial Physics, New Mexico State University, New York University,
Ohio State University, Pennsylvania State University, University of
Portsmouth, Princeton University, the Spanish Participation Group,
University of Tokyo, University of Utah, Vanderbilt University, University
of Virginia, University of Washington, and Yale University. When (some of)
the data reported here were obtained, UKIRT was supported by NASA and
operated under an agreement among the University of Hawaii, the University
of Arizona, and Lockheed Martin Advanced Technology Center; operations were
enabled through the cooperation of the East Asian Observatory.

\end{acknowledgements}


\bibliographystyle{aa}

\onecolumn

\section*{Appendix}
\setcounter{table}{0}
\renewcommand{\thetable}{A\arabic{table}}

\begin{table}[H]
\centering
\rotatebox{90}{%
\resizebox{1.25\linewidth}{!}{%
\begin{tabular}{lllllllllllllllllll}
\toprule
Name & RA & DEC & QSO & BAL & RED & FIRST & Int. Flux [mJy] & Redshift & u & g & r & i & z & W1 & W2 & W3 & W4 & Ref.\\
\midrule
CQ0354-0030 & $58.69183$ & $-0.50819$ & $1$ & $0$ & $1$ & $0$ & $0$ & $1.0$  & $22.0 \pm 0.2$ & $21.32 \pm 0.04$ & $20.33 \pm 0.03$ & $20.03 \pm 0.03$ & $19.59 \pm 0.07$ & $15 \pm 4$ & $14 \pm 4$ & $11.0 \pm 1.0$ & $9 \pm 4$ & F13p \\
cq0206+0624 & $31.5611$ & $6.415122$ & $1$ & $0$ & $1$ & $1$ & $1.57$ & $1.2$ & $21.24 \pm 0.08$ & $19.89 \pm 0.03$ & $19.02 \pm 0.02$ & $18.75 \pm 0.02$ & $18.59 \pm 0.03$ & $15 \pm 4$ & $13 \pm 3$ & $11.0 \pm 0.7$ & $9 \pm 3$ & F13u \\
eHAQ0940+0532 & $145.214$ & $5.537025$ & $1$ & $0$ & $1$ & $0$ & $0$ & $2.32$ & $20.44 \pm 0.06$ & $19.32 \pm 0.02$ & $18.49 \pm 0.01$ & $18.15 \pm 0.03$ & $17.87 \pm 0.03$ & $15 \pm 4$ & $14 \pm 5$ & $11.0 \pm 1.0$ & $8.0 \pm 2.0$ & K16p \\
HAQ2349+0620 & $357.26$ & $6.349694$ & $1$ & $0$ & $1$ & $0$ & $0$ & $2.15$ & $20.28 \pm 0.04$ & $19.6 \pm 0.02$ & $18.81 \pm 0.01$ & $18.53 \pm 0.02$ & $18.13 \pm 0.02$ & $15 \pm 4$ & $14 \pm 5$ & $11.0 \pm 0.9$ & $8.0 \pm 2.0$ & K15p \\
Q564 & $17.1135$ & $28.03839$ & $1$ & $0$ & $0$ & $0$ & $0$ & $3.57$ & $21.1 \pm 0.1$ & $19.21 \pm 0.02$ & $18.03 \pm 0.01$ & $17.91 \pm 0.02$ & $17.66 \pm 0.02$ & $14 \pm 3$ & $14 \pm 4$ & $10.33 \pm 0.05$ & $8.0 \pm 2.0$ & u \\
eHAQ2300+0409 & $345.143202$ & $4.16287468$ & $1$ & $0$ & $0$ & $0$ & $0$ & $3.42$ & $25.0 \pm 1.0$ & $21.0 \pm 5.0$ & $20.0 \pm 3.0$ & $20.0 \pm 3.0$ & $19 \pm 8$ & $15 \pm 5$ & $15.0 \pm 1.30$ & $12.0 \pm 1.0$ & $8.8 \pm 1.0$ & K16u \\
GQ1250+2827 & $192.681029$ & $28.454141$ & $1$ & $0$ & $1$ & $0$ & $0$ & $1.48$ & $22.2 \pm 0.2$ & $20.08 \pm 0.02$ & $19.03 \pm 0.01$ & $18.41 \pm 0.01$ & $18.01 \pm 0.02$ & $14.27 \pm 0.03$ & $13.24 \pm 0.03$ & $10.45 \pm 0.06$ & $8.0 \pm 2.0$ & H20u \\
GaiaQSp1218+0832 & $184.6254$ & $8.537633$ & $1$ & $0$ & $1$ & $0$ & $0$ & $2.6$ & $23.0 \pm 0.4$ & $21.09 \pm 0.04$ & $20.33 \pm 0.03$ & $20.06 \pm 0.03$ & $19.55 \pm 0.06$ & $16.0 \pm 0.7$ & $15.0 \pm 0.9$ & $11.5 \pm 0.2$ & $8 \pm 3$ & G19p \\
GaiaQSp1309+2904 & $197.3497$ & $29.08106$ & $1$ & $1$ & $1$ & $0$ & $0$ & $2.66$ & $22.7 \pm 0.3$ & $21.12 \pm 0.04$ & $19.88 \pm 0.02$ & $19.24 \pm 0.02$ & $18.92 \pm 0.05$ & $15 \pm 4$ & $14 \pm 5$ & $11.0 \pm 0.8$ & $9.0 \pm 2.0$ & F20p \\
GQ124958+273317 & $192.4921$ & $27.95103$ & $1$ & $0$ & $1$ & $0$ & $0$ & $1.95$ & $19.81 \pm 0.04$ & $19.58 \pm 0.02$ & $19.41 \pm 0.02$ & $19.17 \pm 0.03$ & $19.03 \pm 0.04$ & $16.0 \pm 0.7$ & $15.0 \pm 0.8$ & $12.0 \pm 2.0$ & $8.5 \pm 0.3$ & H20p \\
\bottomrule
\end{tabular}%
}}

\begin{minipage}{0.95\textwidth}
\caption{Spectroscopic redshift and astrometric and photometric data of 10 of the 534 total quasars in the eHAQ-GAIA23 sample. The reference column displays both the original sample paper and whether (p) or not (u) the quasar has been published before.}
\label{tab:app}
\end{minipage}
\end{table}

\end{document}